\documentclass[aps,pra, reprint]{revtex4-2}
\usepackage[utf8]{inputenc}
\usepackage{graphicx}
\usepackage{amsmath}
\usepackage{amssymb}
\usepackage{comment}
\usepackage{siunitx}
\usepackage[caption=false]{subfig}
\usepackage{subfiles}
\usepackage{quantikz}
\usepackage{bm}%
\usepackage{array}
\usepackage[version=4]{mhchem}

\DeclareSIUnit{\sq}{\ensuremath{\square}}

\usepackage[colorlinks]{hyperref}
\hypersetup{%
        plainpages=true,
        breaklinks=true,
        hypertexnames=false,
        pageanchor=true,
        colorlinks=true,
        linkcolor={blue},
        citecolor={blue},
        urlcolor={blue},
        anchorcolor={black}
      }

\renewcommand{\eqref}[1]{\mbox{Eq.~(\ref{#1})}}
\newcommand{\tabref}[1]{\mbox{Table~\ref{#1}}}
\newcommand{\figref}[1]{\mbox{Fig.~\ref{#1}}}
\newcommand{\figpanel}[2]{Fig.~\hyperref[#1]{\ref*{#1}(#2)}} 
\newcommand{\figpanels}[3]{Fig.~\hyperref[#1]{\ref*{#1}(#2)-(#3)}} 
\newcommand{\figpanelNoPrefix}[2]{\hyperref[#1]{\ref*{#1}(#2)}} 
\newcommand{\figpanelsNoPrefix}[3]{\hyperref[#1]{\ref*{#1}(#2)-(#3)}} 

\usepackage{mleftright} 

\begin{document}

\title{A quasiparticle-protected superconducting qubit}
\author{Trond Hjerpekjøn Haug}
\email{trond.haug@chalmers.se}
\author{Johan Kolvik}
\author{Joey Frey}
\author{Paul Burger}
\author{Fabian Resare}
\author{Witlef Wieczorek}
\author{Simone Gasparinetti}
\author{Per Delsing}
\author{Rapha\"{e}l Van Laer}
\email{raphael.van.laer@chalmers.se}
\affiliation{Department of Microtechnology and Nanoscience, Chalmers University of Technology, 412 96 Gothenburg, Sweden}
\date{\today}
\begin{abstract}
Superconducting quantum circuits underpin large-scale quantum processors and are increasingly interfaced with optical systems, yet they are vulnerable to quasiparticle poisoning from high-energy radiation and optical fields.
Gap engineering of Josephson junctions suppresses quasiparticle-induced relaxation, but quasiparticles near the junction can still shift the qubit frequency and cause phase errors.
Here, we extend gap engineering beyond the junction using a flux-tunable transmon with a gold-on-niobium ground plane and island and strongly gap-engineered aluminum junctions designed to keep quasiparticles away from the junction leads.
Compared with an all-aluminum transmon with nominally identical junctions, we demonstrate that the niobium device withstands more than an order of magnitude higher near-infrared optical power before excess relaxation appears. 
The Nb transmon shows no resolvable quasiparticle-induced frequency shift over the same power range.
At the same time, we observe that dephasing and state-preparation-and-measurement infidelity increase with optical power even while relaxation remains protected.
These results suggest that engineering the quasiparticle energy landscape beyond the junction can substantially increase qubit resilience, not only to radiation-induced error bursts, but also to nearby optical fields.

\end{abstract}
\maketitle

\textit{Introduction---}
Coherence times in superconducting qubits have increased substantially over the past decades owing to advances in design, fabrication, and packaging~\cite{kochChargeinsensitiveQubitDesign2007,kjaergaardSuperconductingQubitsCurrent2020,balSystematicImprovementsTransmon2024,biznarovaMitigationInterfacialDielectric2024,malevannayaEngineeringGuideSuperconducting2025,tuokkolaMethodsAchieveNearmillisecond2025,blandMillisecondLifetimesCoherence2025}.
These improvements have enabled execution of quantum circuits beyond the reach of classical simulation~\cite{morvanPhaseTransitionsRandom2024,gaoEstablishingNewBenchmark2025_short}, as well as demonstrations of quantum error correction below threshold for the surface code~\cite{googlequantumaiandcollaboratorsQuantumErrorCorrection2025} and color code~\cite{lacroixScalingLogicColour2025}.
In parallel, superconducting circuits are increasingly interfaced with optical systems for microwave-to-optical transduction, quantum networking, control, and readout~\cite{mirhosseiniSuperconductingQubitOptical2020,jiangOpticallyHeraldedMicrowave2023,meesalaNonclassicalMicrowaveOptical2024,meesalaQuantumEntanglementOptical2024}.
These two directions share a common challenge: superconducting circuits are vulnerable to quasiparticles generated by high-energy radiation or optical fields.

In large-scale processors, a high-energy impact on the chip generates a burst of energetic phonons that break Cooper pairs and create quasiparticles (QPs) in the superconducting film.
Quasiparticles that tunnel across a Josephson junction (JJ) can absorb qubit energy and induce relaxation~\cite{martinisEnergyDecaySuperconducting2009,catelaniRelaxationFrequencyShifts2011,sunMeasurementsQuasiparticleTunneling2012}.
Although such impacts are rare on the timescale of a single physical-qubit lifetime, the excited QPs can persist and affect many qubits simultaneously.
The resulting correlated error bursts are not correctable with standard quantum error-correcting codes~\cite{fowlerQuantifyingEffectsLocal2014,kamDetrimentalNonMarkovianErrors2025}.

One approach to mitigate high-energy impacts is to make qubits resilient to QP poisoning.
Engineering different superconducting gaps across a transmon's JJs can strongly suppress QP-induced relaxation~\cite{aumentadoNonequilibriumQuasiparticles22004,fergusonMicrosecondResolutionQuasiparticle2006,yamamotoParityEffectSuperconducting2006,courtEnergyGapMeasurement2008,risteMillisecondChargeparityFluctuations2013,riwarEfficientQuasiparticleTraps2019,diamondDistinguishingParitySwitchingMechanisms2022,panEngineeringSuperconductingQubits2022,marchegianiQuasiparticlesSuperconductingQubits2022,mcewenResistingHighenergyImpact2024,nhoRecoveryDynamicsGapEngineered2026}.
When the gap difference exceeds the qubit energy, QPs near the gap edge cannot tunnel into the higher-gap electrode by absorbing energy from the qubit.
Gap engineering, however, suppresses QP tunneling rather than the QP population itself.
Quasiparticles near the JJ still reduce the supercurrent and shift the qubit frequency.
Following a high-energy impact, the time-dependent QP density can therefore produce frequency fluctuations and dephasing.
Junction gap engineering can thus protect against QP-induced relaxation while leaving the qubit susceptible to QP-induced phase errors~\cite{kurilovichCorrelatedPhaseError2026}.

The same QP challenge arises when superconducting circuits are intentionally operated near optical fields.
A prominent example is microwave-to-optical transduction, where superconducting circuits are coupled to optically pumped transducers~\cite{mirhosseiniSuperconductingQubitOptical2020,jiangOpticallyHeraldedMicrowave2023,meesalaNonclassicalMicrowaveOptical2024,meesalaQuantumEntanglementOptical2024,burgerDesignReleasefreePiezooptomechanical2025,burgerReleasefreeElectrooptomechanicalCrystal2026}.
Related efforts increasingly use optical fields for the control and readout of superconducting systems.
Stray infrared photons from these optical fields have energies far above the superconducting gap and can therefore break Cooper pairs and generate QPs.
Suppressing QP poisoning is thus a key requirement for superconducting circuits that must operate in proximity to optical fields.

In this work, we extend gap engineering beyond the JJ by engineering the quasiparticle energy landscape of the transmon to keep QPs away from the junction leads.
Our device combines a Nb/Au ground plane and qubit island with strongly gap-engineered Al JJs, and we compare its optical response with an all-Al transmon designed to reproduce the strongly gap-engineered devices of Ref.~\cite{mcewenResistingHighenergyImpact2024}.
Under continuous $\SI{1550}{\nano\meter}$ illumination, the Nb transmon withstands more than an order of magnitude higher optical power before excess energy relaxation appears.
Over the same power range, we resolve no QP-induced frequency shift, consistent with a low QP density near the JJs.
Dephasing and state-preparation-and-measurement (SPAM) infidelity nevertheless increase with optical power even while the relaxation rate remains unchanged, revealing additional error channels once QP-induced relaxation is suppressed.
Together, these results show that engineering the quasiparticle energy landscape beyond the junction can substantially increase qubit resilience while identifying remaining limitations relevant to both radiation-induced error bursts and superconducting circuits operated near optical fields.

\textit{Experimental design}---
The transmon qubits are fabricated with nominally identical Manhattan-style Al/AlO$_x$/Al JJs.
We engineer the superconducting gap $\Delta$ of each lead through its film thickness $t$, targeting
$t_\mathrm{thick} = \SI{100}{\nano\meter}$ for the thicker, low-gap lead and
$t_\mathrm{thin} = \SI{18}{\nano\meter}$ for the thinner, high-gap lead.
Using the empirical model of Refs.~\cite{marchegianiQuasiparticlesSuperconductingQubits2022,mcewenResistingHighenergyImpact2024}, we estimate a gap difference $\delta\Delta \approx \SI{40}{\micro\eV}$, corresponding to $\delta\Delta/h \approx \SI{10}{\giga\hertz}$.
Two nominally identical junctions form a SQUID, giving flux-tunable qubits with maximum frequencies near $\SI{5}{\giga\hertz}$.
The engineered gap difference therefore exceeds the maximum frequency for both transmons.
We avoid Al films thinner than $\SI{18}{\nano\meter}$ because of fabrication reproducibility and the need for the junction leads to withstand ion milling.

We use an all-Al chip as the reference device.
Its ground plane and qubit island are formed from $\SI{300}{\nano\meter}$ of evaporated Al on high-resistivity silicon.
The transmon is designed to reproduce the strongly gap-engineered qubits of Ref.~\cite{mcewenResistingHighenergyImpact2024}, with each JJ connected to the qubit island and ground plane by $\SI{300}{\nano\meter}$-thick Al bandages.
The same chip also contains a high-internal-quality-factor Al resonator, which serves as a reference for optically induced, distributed microwave loss in the Al film.

Our light-resilient transmon is fabricated on a separate chip with a $\SI{150}{\nano\meter}$ sputtered Nb ground plane and qubit island.
Niobium is attractive for radiation-resilient circuits because nonequilibrium QPs can relax on nanosecond timescales~\cite{warburtonQuasiparticleLifetimesEpitaxial1995,loboPhotoinducedTimeresolvedElectrodynamics2005}, much faster than the millisecond-scale QP lifetimes reported in Al~\cite{devisserNumberFluctuationsSparse2011}.
Moreover, the large superconducting gap of Nb reduces the cross section for secondary QP generation events from high-energy phonons.
We cap the Nb in situ with $\SI{10}{\nano\meter}$ of Au to suppress formation of the lossy native Nb oxide, an approach shown to improve qubit relaxation times~\cite{balSystematicImprovementsTransmon2024}. We expect the thin Au layer to be strongly proximitized by the underlying Nb and to leave the Nb gap largely unchanged.
At the same time, thin Au films on Nb can host subgap states~\cite{guptaAnomalousDensityStates2004}, which could improve resilience by providing additional QP trapping.
We expect the reflectance of Al and Nb+Au films at $\SI{1550}{\nano\meter}$ to be similar. 

\begin{figure}
    \centering
    \includegraphics[width=\linewidth]{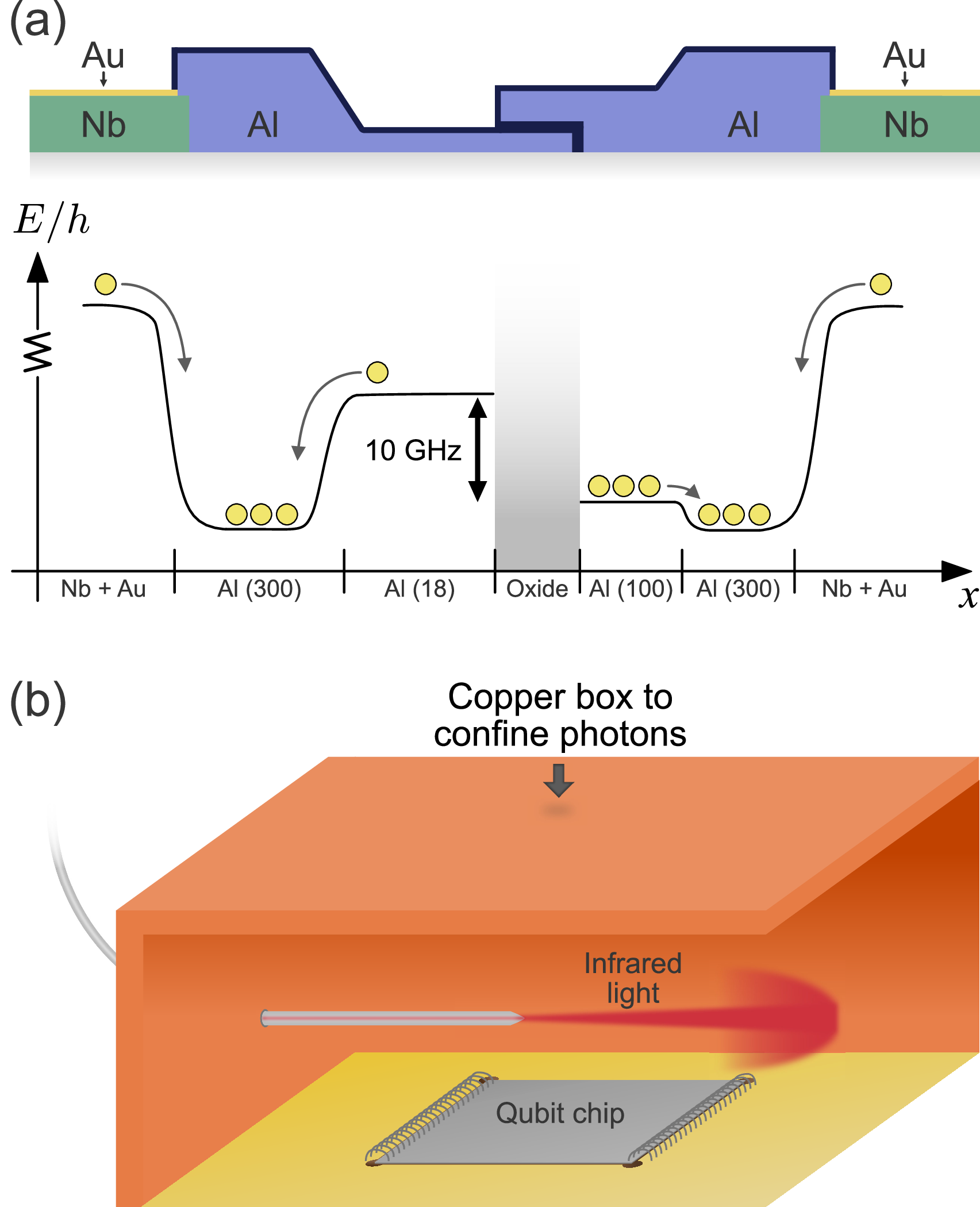}
    \caption{\textbf{Device design and optical-illumination setup.}
(a) Schematic of the materials near a transmon Josephson junction (top) and the corresponding QP energy landscape (bottom).
Film thicknesses are not to scale.
Al and AlO$_x$ are shown in light and dark blue, respectively, with Al thicknesses in nm indicated in parentheses.
Quasiparticles are depicted as yellow spheres and are expected to preferentially relax into the lowest-gap Al regions, here the thick Al bandages away from the tunnel barrier.
(b) Optical-illumination setup.
Infrared light is guided through an optical fiber into the copper package containing the qubit chip.
Reflections from the copper enclosure distribute the light across the chip to promote spatially distributed QP generation.}
    \label{fig:setup}
\end{figure}

The Nb transmon is designed to create a QP energy landscape that draws QPs into low-gap regions away from the JJs.
Quasiparticles generated in Nb ($\Delta/h \sim \SI{360}{\giga\hertz}$) are expected to recombine quickly, but those that are generated close to the junctions can diffuse into the lower-gap Al ($\Delta/h \sim \SI{50}{\giga\hertz}$), where they can become trapped as they relax towards the Al gap edge.
To keep QPs away from the tunnel barriers, the JJ leads are connected to the surrounding Nb through $\SI{300}{\nano\meter}$-thick Al bandages whose gap is lower than that of the junction leads.
Before evaporating the bandages, we remove the Au at the contact regions by ion milling to ensure direct Nb--Al contact.
The resulting QP energy landscape near a junction is illustrated in \figpanel{fig:setup}{a}.

The chips are mounted in oxygen-free copper sample boxes attached to the mixing-chamber plate of a dilution refrigerator.
A $\SI{1550}{\nano\meter}$ laser is coupled into each package through an optical fiber with a cleaved tip inserted through the sample-box wall.
The fiber tip is positioned $\sim\SI{5}{\milli\meter}$ above the chip center and intentionally directed away from the chip, as shown in \figpanel{fig:setup}{b}.
A DC-driven coil above the package provides the magnetic field used to tune the qubit frequency.
Further details of the measurement setup are given in Appendix~\ref{sec:setup}.

\begin{figure*}
    \centering
    \includegraphics[width=\linewidth]{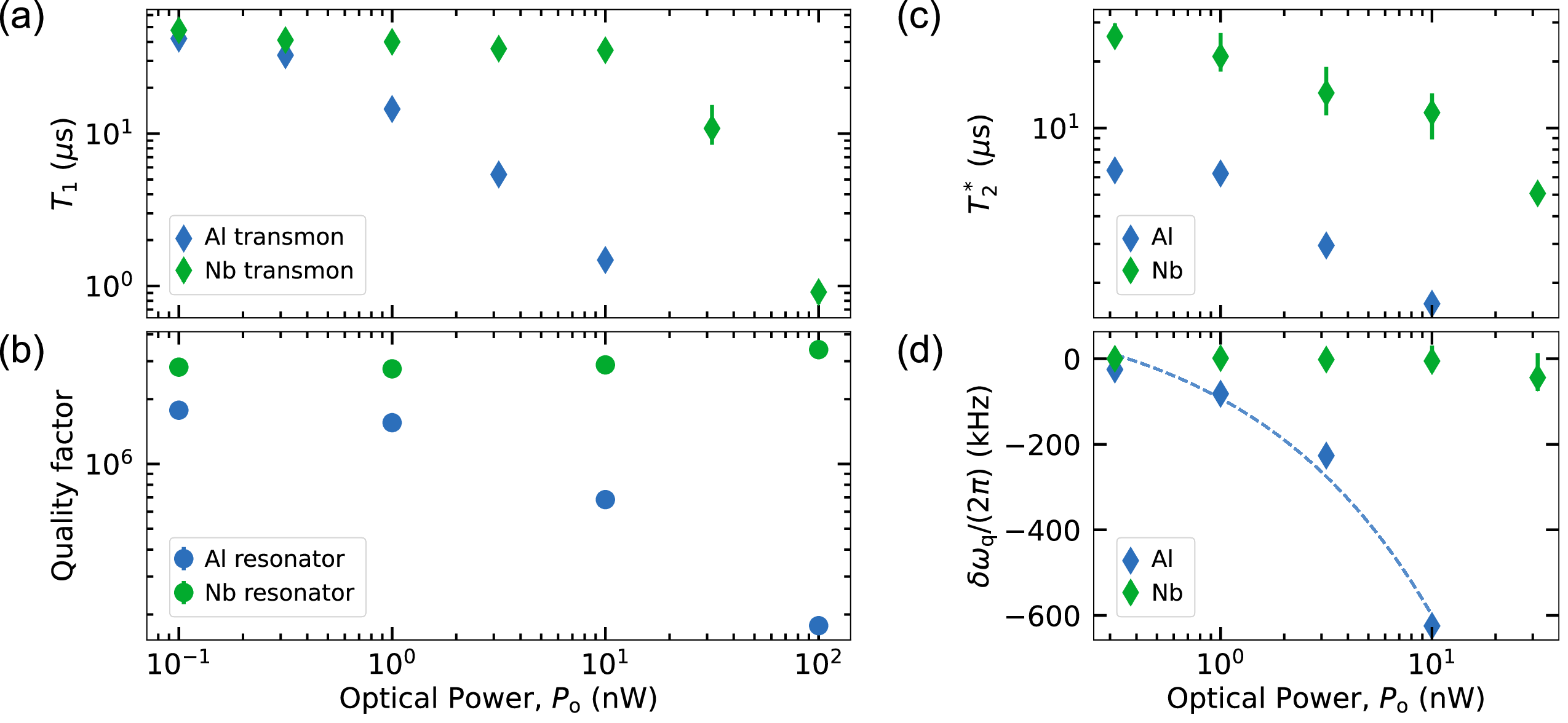}
    \caption{\textbf{Coherence and frequency response under optical illumination.}
    (a) $T_1$ times of the Al and Nb transmons and the Al resonator versus optical power at the fiber tip.
    Qubit $T_1$ values are extracted from exponential fits to population decay traces averaged over at least 2000 repetitions.
    For the Nb transmon, vertical bars indicate the spread across three series of measurements performed on different days.
    (b) Internal quality factor for resonators made from Al and Nb. 
    (c) Ramsey coherence time $T_2^*$ versus optical power, with both qubits operated at their maximum frequency.
    (d) Qubit frequency shift extracted from Ramsey measurements.
    The dashed curve shows a fit $\delta\omega_q\propto\sqrt{P_\mathrm{o}}$, expected when the frequency shift is proportional to QP density and the steady-state QP density is limited by recombination~\cite{mcewenResistingHighenergyImpact2024}.}
    \label{fig:coherence}
\end{figure*}

\textit{Results}---
We first measure the qubit relaxation time $T_1$ as a function of continuous optical power $P_\mathrm{o}$.
The Al and Nb transmons are operated at their maximum frequencies of approximately $\SI{5.4}{\giga\hertz}$ and $\SI{4.4}{\giga\hertz}$, respectively.
We use the onset of excess relaxation as our primary measure of resilience to QP poisoning, since QP tunneling across a JJ can induce qubit decay.
\figpanel{fig:coherence}{a} shows $T_1$ for the Al and Nb transmons.
For $P_\mathrm{o} \leq \SI{0.1}{\nano\watt}$, both devices have $T_1 \sim \qtyrange{40}{50}{\micro\second}$, with no measurable optically induced excess relaxation.
Above this power, the Al transmon shows a clear increase in relaxation rate, consistent with QP poisoning and closely matching the behavior reported in Ref.~\cite{mcewenResistingHighenergyImpact2024}.
By contrast, the Nb transmon shows no measurable increase in $\Gamma$ for $P_\mathrm{o} \leq \SI{10}{\nano\watt}$.
Above $\SI{10}{\nano\watt}$, $\Gamma$ increases sharply, and the qubit response vanishes for $P_\mathrm{o} > \SI{100}{\nano\watt}$.

Quasiparticle-induced relaxation can depend on the qubit frequency.
However, the frequencies for both the Al and Nb qubits remain substantially below the estimated gap difference of $\delta\Delta/h \approx \SI{10}{\giga\hertz}$. 
Moreover, previous optical-illumination measurements in Ref.~\cite{mcewenResistingHighenergyImpact2024} found no significant difference in robustness among strongly gap-engineered qubits operated at different frequencies within the same protected regime.
The frequency difference may contribute to the quantitative comparison, with the higher-frequency Al transmon expected to be somewhat more sensitive, but cannot account for the large observed difference in the onset of QP-induced relaxation. 

Next, we measure the relaxation rate of the high-quality resonator on the reference Al chip and a separate Nb chip as a function of optical power.
The loss in the Al resonator increases monotonically with $P_\mathrm{o}$, as shown in \figpanel{fig:coherence}{b}, consistent with increasing distributed microwave loss from optically generated QPs.
On the other hand, the resonator on a Nb chip shows no measurable linewidth increase for $P_\mathrm{o} \leq \SI{10}{\micro\watt}$.
In fact, the internal quality factor of the Nb resonator appears to increase slightly with increasing $P_\mathrm{o}$, reaching $\sim\SI{20}{\percent}$ higher than its zero-power value at $P_\mathrm{o} = \SI{100}{\nano\watt}$. 

\begin{figure*}
    \centering
    \includegraphics[width=\linewidth]{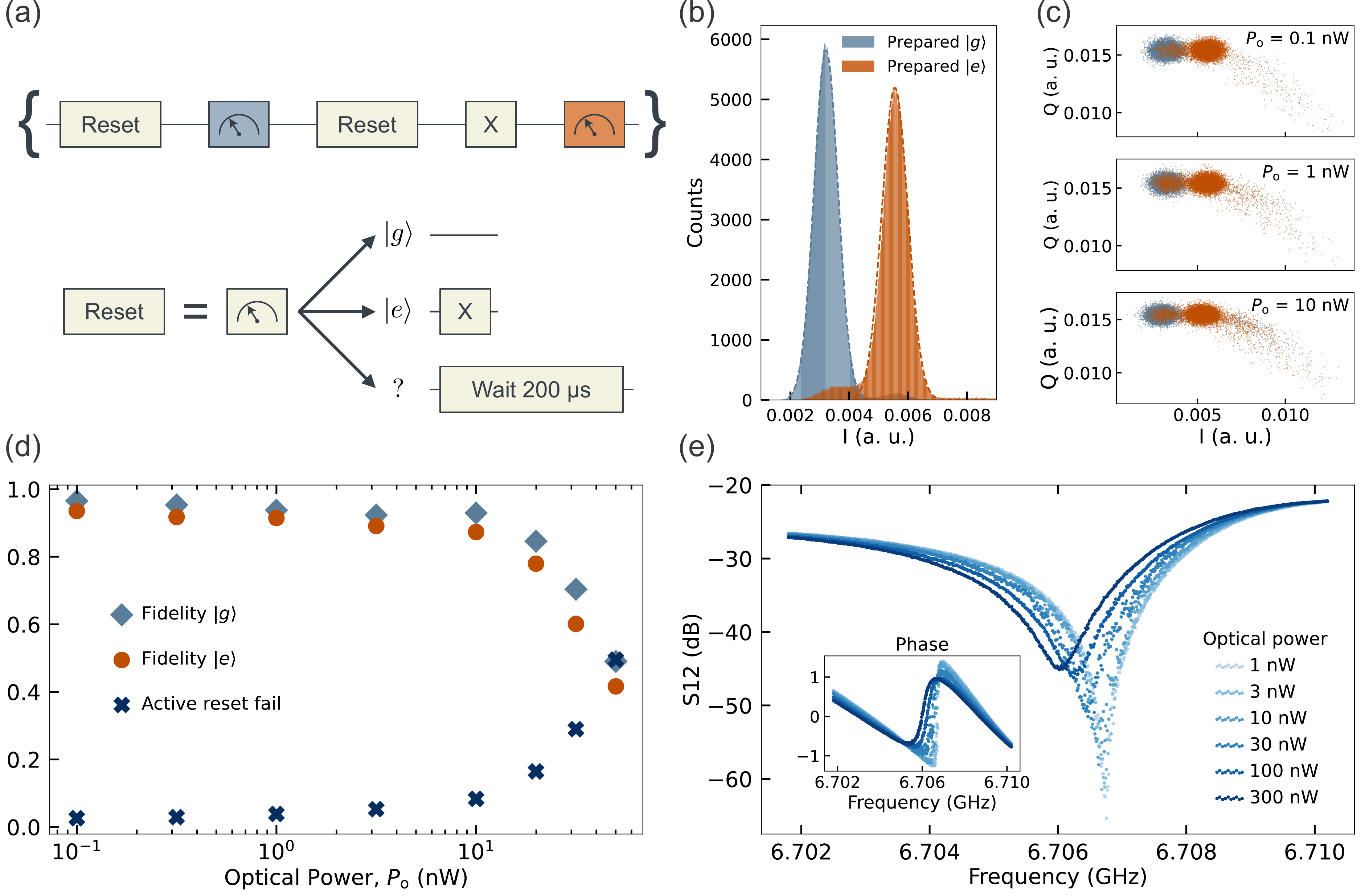}
    \caption{\textbf{Qubit readout under optical illumination.}
    (a) Protocol for state-preparation and measurement (top) and active reset (bottom). 
    The protocol in the curly brackets is repeated $10^6$ times without gaps between iterations.
    (b) Readout distributions following reset and preparation in $\vert g\rangle$ or $\vert e\rangle$.
    Dashed lines represent Gaussian fits to the distributions. 
    (c) Distribution of the first $10^4$ shots in the IQ plane for three different levels of optical power.
    The long tail at large I-quadrature values is attributed to transmon states outside the qubit manifold.
    (d) SPAM fidelity and fraction of resets triggered by out-of-manifold outcomes.
    (e) VNA response of the readout resonator coupled to the Nb transmon.}
    \label{fig:readout}
\end{figure*}

The onset of degradation in the Nb transmon at high optical power could be caused by QPs tunneling across the JJs despite the engineered gap difference, for example after QPs absorb high-frequency phonons generated by recombination in the Nb film~\cite{connollyCoexistenceNonequilibriumDensity2024}.
We discuss this possibility in Appendix~\ref{sec:hot_QPs}.
An alternative explanation for the observed increase in qubit relaxation at high optical power may be found in the transmon's Al components.
In particular, the transmon contains $\SI{300}{\nano\meter}$-thick, low-gap Al bandages designed to trap QPs.
A sufficiently large QP population in these bandages would produce dissipative microwave loss and could reduce the transmon $T_1$.
In Appendices~\ref{sec:QP_density_res}--\ref{sec:simulations}, we estimate the QP-induced resistance required to reproduce the measured transmon loss.
Within this model, the bandage resistance required to account for the data is approximately $500$ times larger than that inferred for the Al resonator at comparable optical power, corresponding to a similarly enhanced QP density under the model assumptions.
We cannot determine from these measurements whether such a QP density is actually reached in the bandages.

The qubit transition frequency $\omega_\mathrm{q}$ is sensitive to the Josephson energy
$E_\mathrm{J} = \Phi_0 I_c/2\pi$,
where $I_\mathrm{c}$ is the critical current and $\Phi_0 = h/2e$ is the flux quantum given by the Planck constant $h$ and the elementary charge $e$.
Quasiparticles in the JJ leads suppress the critical current and, consequently, the Josephson energy, resulting in a downward shift of the qubit frequency.
The induced frequency shift $\delta\omega_\mathrm{q} = \omega_\mathrm{q}(P_\mathrm{o}) - \omega_\mathrm{q}(0)$ therefore provides a probe of the QP population near the junctions.
By contrast, QPs trapped in the Al bandages are spatially separated from the tunnel barriers and are expected to have a much weaker effect on the Josephson energy.
If QPs preferentially accumulate in the bandages rather than the JJ leads, we expect a reduced QP-induced frequency shift.

We use Ramsey measurements to extract $T_2^*$ and $\delta\omega_\mathrm{q}$ as a function of optical power, as shown in \figpanels{fig:coherence}{c}{d}.
The Al transmon exhibits an optical-power-dependent frequency shift similar to that reported in Ref.~\cite{mcewenResistingHighenergyImpact2024}.
By contrast, we resolve no frequency shift in the Nb transmon for $P_\mathrm{o} \leq \SI{10}{\nano\watt}$, consistent with a low QP population near the JJs.
Above $\SI{10}{\nano\watt}$, decreasing measurement fidelity limits the reliable extraction of $T_2^*$ and $\delta\omega_\mathrm{q}$, as discussed below.
Unlike $T_1$, however, the Nb-transmon $T_2^*$ degrades with optical power already below $\SI{10}{\nano\watt}$.
Even at zero optical power, the $T_2^*$ of the Nb transmon is approximately five times that of the Al transmon. The origin of this difference is unclear; typical $T_2^*$ values for all flux-tunable qubits in our setup with no optical illumination are below $\SI{10}{\micro\second}$.

Next, we characterize the fidelity of state preparation and measurement (SPAM) of the Nb transmon under optical illumination.
The protocol is described in \figpanel{fig:readout}{a}. 
Each reset cycle begins with a measurement, followed by a conditional $\pi^{ge}$ pulse if the transmon is found in $\vert e\rangle$.
Outcomes inconsistent with either $\vert g\rangle$ or $\vert e\rangle$ instead trigger a $\SI{200}{\micro\second}$ passive reset.
The fraction of such events provides a measure of the probability that the transmon occupies a state outside the qubit manifold.
We refer to excursions outside the qubit manifold before or during measurement as \textit{ionization}~\cite{sankMeasurementInducedStateTransitions2016,khezriMeasurementinducedStateTransitions2023,dumasMeasurementInducedTransmonIonization2024}.
After reset, we prepare the transmon in either $\vert g\rangle$ or $\vert e\rangle$ and measure it again.
Readout uses a $\SI{2}{\micro\second}$ square pulse centered on the resonator frequency.
The readout frequency and power have been chosen to maximize the signal-to-noise ratio while limiting ionization events below $\SI{2}{\percent}$ in the absence of optical illumination. 

Histograms from $10^6$ repetitions of the reset protocol at $P_\mathrm{o} = \SI{0.1}{\nano\watt}$ are shown in \figpanel{fig:readout}{b}.
The transmitted-signal I quadrature clearly discriminates the prepared $\vert g\rangle$ and $\vert e\rangle$ states, with an overlap error of $\SI{0.3}{\percent}$ calculated by fitting each distribution to a Gaussian. 
Both distributions also exhibit a long tail inconsistent with either state, visible in the IQ plane for around 1\% of the measurements in \figpanel{fig:readout}{c}.
Most events in this tail are also inconsistent with occupation of $\vert f\rangle$.
To test this, we apply consecutive $\pi^{fe}$ and $\pi^{eg}$ pulses before measurement and observe no increase in the ground-state population beyond that expected from energy relaxation.
Only passive thermalization reliably returns these events to $\vert g\rangle$.
We therefore interpret the tail as ionization of the transmon induced by the combination of optical illumination and strong microwave readout.
Ionization occurs for both initially prepared qubit states.

We repeat the reset protocol as a function of optical power and plot the SPAM fidelity together with the fraction of unsuccessful active resets triggered by out-of-manifold outcomes in \figpanel{fig:readout}{d}.
Two regimes emerge.
For $P_\mathrm{o} \lesssim \SI{10}{\nano\watt}$, the SPAM fidelity decreases gradually with optical power, while above this level it degrades rapidly.
The fraction of passive resets shows a similar crossover near $\SI{10}{\nano\watt}$, coincident with the onset of $T_1$ degradation in \figpanel{fig:coherence}{a}.
Below this threshold, preparation and measurement of $\vert g\rangle$ and $\vert e\rangle$ degrade similarly despite the absence of measurable excess relaxation.
For $P_\mathrm{o} > \SI{30}{\nano\watt}$, the SPAM fidelity becomes too low to reliably extract $T_2^*$ and $\delta\omega_\mathrm{q}$ from Ramsey measurements.

Optical illumination also modifies the response of the readout resonator coupled to the Nb transmon, as shown in \figpanel{fig:readout}{e}.
With increasing $P_\mathrm{o}$, the resonance shifts toward the bare-resonator frequency while its linewidth broadens.
A possible explanation is that the combination of optical illumination and strong microwave readout drives the transmon into highly excited states whose populations fluctuate in time.
Because the dispersive shift depends on the transmon state, such fluctuations would produce time-dependent resonator-frequency shifts, leading to spectral broadening and reduced notch depth.
A similar readout-resonator response under pulsed optical illumination has been reported for a dispersively coupled transmon~\cite{liSuperfluidCooledTransmonQubits2025}.
The bare Nb readout resonator, when measured in the high-power regime where its response is no longer dressed by the transmon, shows no measurable response to optical power.

Finally, we use Rabi population measurements~\cite{geerlingsDemonstratingDrivenReset2013} to extract the residual excited-state population $p_\mathrm{e}$ after thermal reset.
We find $p_\mathrm{e} \in (0.01,0.02)$ for $P_\mathrm{o} \leq \SI{30}{\nano\watt}$, corresponding to an effective transmon temperature of approximately $\SI{50}{\milli\kelvin}$.
The residual excited-state population shows no systematic increase with $P_\mathrm{o}$, consistent with previous measurements of optically illuminated transmons~\cite{benevidesQuasiparticleDynamicsSuperconducting2024}.
Further details are given in Appendix~\ref{sec:thermometry}.

\textit{Conclusion and outlook---}
We demonstrate a transmon architecture that extends gap engineering beyond the JJ by engineering the landscape of the superconducting gap across the entire device.
Combining a Nb/Au ground plane and island with strongly gap-engineered Al JJs increases the optical power required to induce excess relaxation by more than an order of magnitude compared with an all-Al transmon with nominally identical junctions.
Over the same power range, we resolve no QP-induced frequency shift, consistent with a low QP density near the junction leads.
Together, these results show that engineering the gap landscape beyond the junction can suppress multiple effects of QP poisoning.

The protection is not complete, however.
Dephasing and state-preparation-and-measurement (SPAM) infidelity increase with optical power even while the relaxation rate remains unchanged, revealing additional error channels once QP-induced relaxation is suppressed.
Further improvements will require identifying these remaining mechanisms and controlling QP generation, transport, and trapping throughout the device.
Promising directions include reducing the role of Al in the most QP-sensitive regions, exploring alternative junction materials, and understanding whether the Au encapsulation can be engineered to provide additional QP trapping.
Together with improved suppression of stray optical light, these approaches could substantially extend the optical-power range accessible to superconducting qubits and facilitate their integration with optically driven quantum devices.
More broadly, device-level control of quasiparticles provides a route toward superconducting qubits that better withstand both radiation-induced error bursts and nearby optical fields.

\textit{Acknowledgments}---
We thank Amr Osman, Claudia Castillo-Moreno, Kunal Helambe, and Simon Sundelin for advice that enabled the fabrication of the samples.
We thank Hampus Renberg Nilsson for providing the TWPA used in the experiments. 
The devices were fabricated in the MyFab Nanofabrication Laboratory at Chalmers.
W.W. acknowledges support in part by the Knut och Alice Wallenberg Foundation through a Wallenberg Scholar grant.
R.V.L. acknowledges support from the Knut and Alice Wallenberg Foundation through the Wallenberg Centre for Quantum Technology through a WACQT Fellowship, from the European Research Council via Starting Grant 948265, and from the Swedish Foundation for Strategic Research via grant FFL21-0039.
S.G. acknowledges financial support from the European Research Council (Grant No. 101041744 ESQuAT), and from the Knut and Alice Wallenberg Foundation through the Wallenberg Centre for Quantum Technology (WACQT).
External interest disclosure: S.G. is a co-founder and equity holder in Sweden Quantum AB.

\appendix

\section{Device design and parameters}
We target a maximum qubit frequency of $\SI{5.1}{\giga\hertz}$ for both the Nb and Al transmons by giving them identical capacitance to ground ($\SI{90}{\femto\farad}$ in simulation) and nominally the same junction area of $\SI{130}{\nano\meter} \times \SI{130}{\nano\meter}$. 
The Al base layer is etched isotropically with a wet etch that removes $\sim\SI{600}{\nano\meter}$ from the linewidth, whereas the Nb base layer is etched anisotropically with a Cl-based plasma dry etch with negligible linewidth loss. 
The fabrication of the Al chip was performed twelve months before the fabrication of the Nb chip, which could introduce process variations leading to deviations in the maximum frequencies of the two transmons. 
To estimate the superconducting gap in the JJs, we use the empirical model $\Delta (t) \propto A/t$ with $A = \SI{900}{\micro\eV \cdot \nano\meter}$ from Refs.~\cite{marchegianiQuasiparticlesSuperconductingQubits2022, mcewenResistingHighenergyImpact2024}.

\begin{figure}
    \centering
    \includegraphics[width=\linewidth]{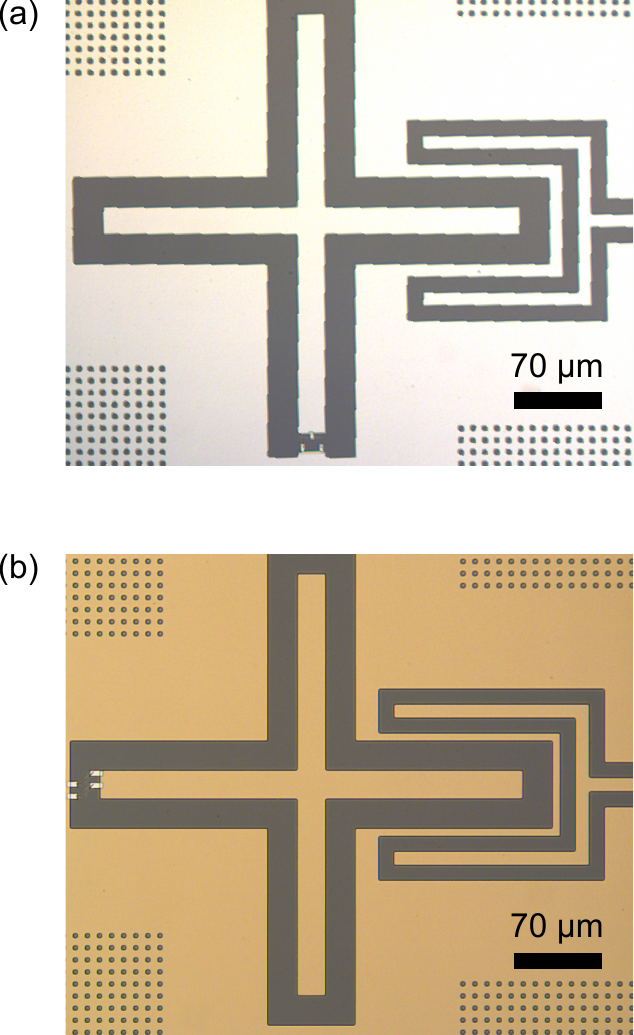}
    \caption{Optical micrographs of (a) the Al transmon and (b) the Nb transmon in this study.}
    \label{fig:images}
\end{figure}

We summarize the relevant parameters for the two transmon qubits in this study in \tabref{tab:parameters}. Optical micrographs of the transmon qubits are shown in \figref{fig:images}. 
\begin{table}[b]
    \centering
    \setlength{\tabcolsep}{3pt}
    \begin{tabular}{lccc}
    \toprule
        \textbf{Parameter} & \textbf{Symbol} & \textbf{Nb} & \textbf{Al}\\
        Maximum freq. & $\omega_\mathrm{q}/2\pi$ & $\SI{4.404}{\giga\hertz}$ & $\SI{5.389}{\giga\hertz}$ \\
        Anharmonicity & $E_C /h$ & $-\SI{224}{\mega\hertz}$ & $-\SI{216}{\mega\hertz}$ \\
        Readout res. freq. & $\omega_\mathrm{r}/2\pi$ & $\SI{6.705}{\giga\hertz}$ & $\SI{6.624}{\giga\hertz}$ \\
        Qubit lifetime & $T_1$ & 40--50 $\SI{}{\micro\second}$ & 40--50 $\SI{}{\micro\second}$ \\
        Ramsey coherence time & $T_2^*$ & 20--30 $\SI{}{\micro\second}$ & 5--6 $\SI{}{\micro\second}$ \\
        Thin-lead thickness & $t_\mathrm{thin}$ & $\SI{18}{\nano\meter}$ & $\SI{18}{\nano\meter}$ \\
        Thick-lead thickness & $t_\mathrm{thick}$ & $\SI{100}{\nano\meter}$ & $\SI{100}{\nano\meter}$ \\
        Bandage thickness & $t_\mathrm{bandage}$ & $\SI{300}{\nano\meter}$ & $\SI{300}{\nano\meter}$ \\
    \end{tabular}
    \caption{Device parameters for the transmon qubits without laser light.}
    \label{tab:parameters}
\end{table}

\section{Measurement setup}
\label{sec:setup}
\begin{figure*}
    \centering
    \includegraphics[width=\linewidth]{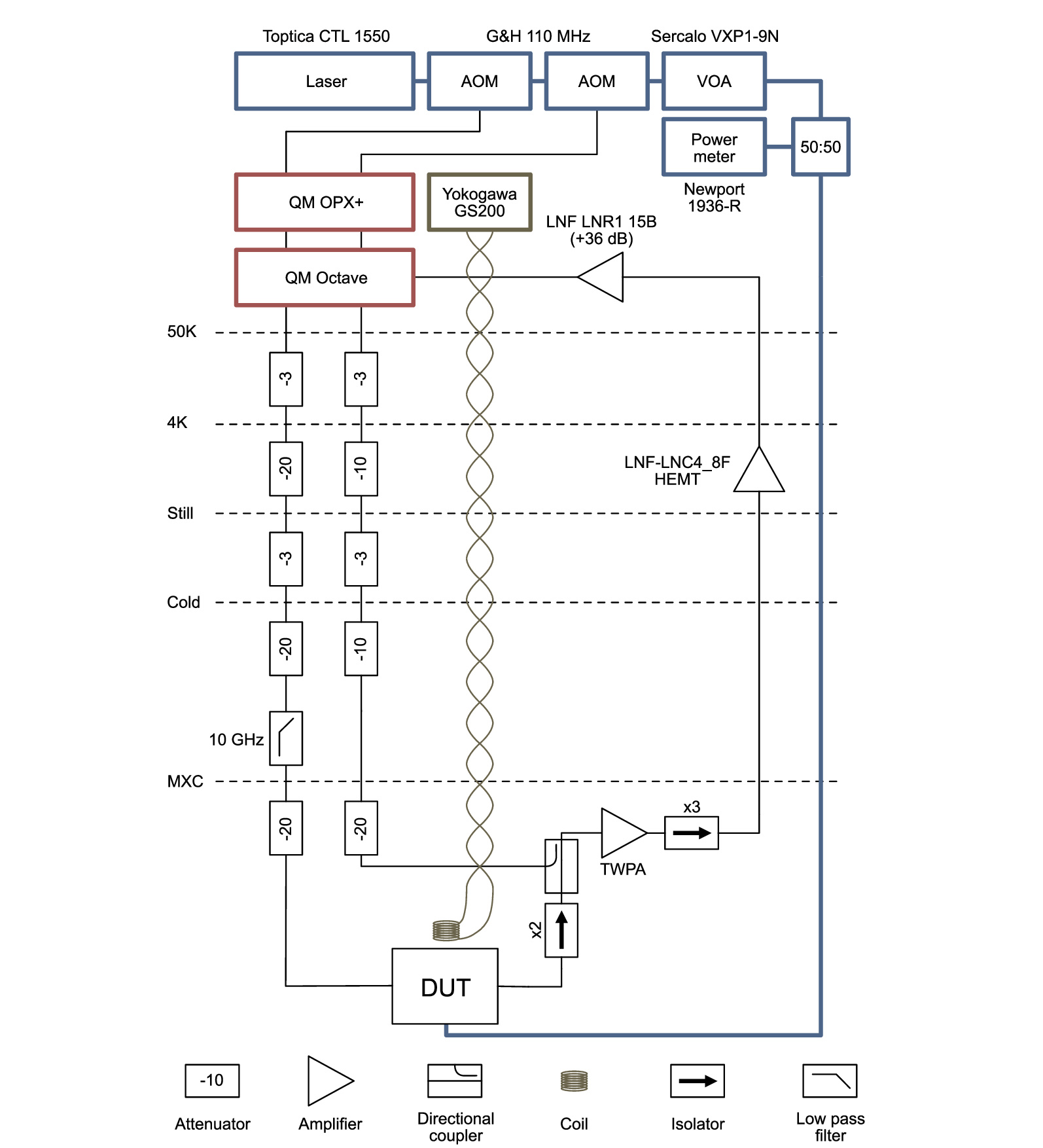}
    \caption{Wiring diagram of the measurement setup. Baseband signals are generated by the OPX+ controller from Quantum Machines and sent to the Octave instrument hosting the LO and IQ mixers for up-conversion to microwave signals.
    The OPX+ is also used to control the acousto-optic modulators (AOMs) which generate optical pulses by periodically blocking the laser.
    The laser power is controlled by a variable optical attenuator from a separate computer.
    Signals from the device under test (DUT) are amplified by a traveling-wave parametric amplifier (TWPA) and a high-electron-mobility transistor (HEMT) low-noise amplifier before exiting the cryostat. 
    Numbers on input lines represent attenuation in dB.}
    \label{fig:measurement_chain}
\end{figure*}

After fabrication, the transmon chip is wire-bonded to a printed circuit board inside the copper package and mounted to the mixing chamber of a Bluefors LD400 dilution refrigerator. 
We place a $\mu$-metal shield around the copper package to protect our flux-tunable transmons from stray magnetic fields. 
The base temperature at the mixing chamber remains at $\sim\SI{7}{\milli\kelvin}$ for all levels of optical power in this study.
A wiring diagram of the system used for measurements is shown in \figref{fig:measurement_chain}.

The optical power at the chip is controlled using a variable optical attenuator (VOA) followed by a power meter. 
Before we cool down the chip, we measure \SI{1.28}{\decibel} optical loss between the 50:50 beam splitter and the fiber tip inside the copper package. 
We also measure the power ratio between the output ports from the splitter and find \SI{0.55}{\decibel} difference, giving a power discrepancy between the power meter and the fiber tip of \SI{1.83}{\decibel}. 
All optical powers in this study refer to the power at the fiber tip. 
We adjust the VOA so that the measured optical power deviates less than \SI{0.5}{\decibel} from the target value for all measurements. 

\section{Quasiparticle density in the Al resonator}
\label{sec:QP_density_res}
Here, we present calculations of the QP density in the $\SI{300}{\nano\meter}$ Al film of resonator using the measured quality factor $Q$.
The frequency-dependent complex conductivity $\sigma(\omega) = \sigma_1(\omega) + i\sigma_2(\omega)$ of a superconducting film can be described using Mattis-Bardeen theory~\cite{mattisTheoryAnomalousSkin1958}. 
When combined with BCS theory for quasiparticles, Gao et al.~\cite{gaoEquivalenceEffectsComplex2008} showed that, under the condition that $\hbar\omega, k_\mathrm{B} T \ll \Delta$, the QP density in the superconductor is
\begin{equation}
    n_\mathrm{qp} = 2N_0\sqrt{2\pi k_\mathrm{B} T \Delta}\exp\left({-\frac{\Delta - \mu}{k_\mathrm{B} T}}\right),
    \label{eq:n}
\end{equation}
and the conductivity is 
\begin{equation}
    \frac{\sigma_1}{\sigma_n} = \frac{4\Delta}{\hbar\omega}\exp\left({-\frac{\Delta - \mu}{k_\mathrm{B} T}}\right)\sinh\left({\frac{\hbar\omega}{2k_\mathrm{B}T}}\right)K_0\left({\frac{\hbar\omega}{2k_\mathrm{B}T}}\right),
    \label{eq:sigma_1}
\end{equation}
\begin{equation}
    \frac{\sigma_2}{\sigma_n} = \frac{\pi\Delta}{\hbar\omega}\left(1-2\exp\left[ -\frac{\Delta - \mu + \hbar\omega/2}{k_\mathrm{B} T}\right] I_0\left[{\frac{\hbar\omega}{2k_\mathrm{B}T}}\right]\right),
\end{equation}
where $\sigma_n$ is the normal-state conductivity, $N_0$ is the single spin density of states, $k_\mathrm{B}$ is the Boltzmann constant, $T$ is the temperature, $\Delta$ is the superconducting gap parameter, $\mu$ is an effective chemical potential, and $I_0, K_0$ are the zeroth-order modified Bessel functions of the first and second kinds, respectively.
Inserting \eqref{eq:n} into \eqref{eq:sigma_1} and using $\hbar\omega \gg k_\mathrm{B}T$, we obtain 
\begin{equation}
    \frac{\sigma_1}{\sigma_n} \approx \frac{1}{2}\frac{n_\mathrm{qp}}{2N_0\Delta} \left(\frac{2\Delta}{\hbar\omega}\right)^\frac{3}{2},
    \label{eq:sigma_1_simplified}
\end{equation}
\begin{equation}
    \frac{\sigma_2}{\sigma_n} \approx \frac{\pi \Delta}{\hbar\omega},
    \label{eq:sigma_2_simplified}
\end{equation}
where $n_\mathrm{qp}/2N_0\Delta \equiv x_\mathrm{qp}$ is the QP density normalized by the number of Cooper pairs.

We can use the link between $\sigma_1$ and $x_\mathrm{qp}$ in \eqref{eq:sigma_1_simplified} to estimate the QP density required to explain the degradation of the Al resonator's quality factor [\figpanel{fig:coherence}{b}]. 
For a superconductor with $x_\mathrm{qp} \ll 1$, we have $\sigma_1 \ll \sigma_2$ and therefore the effective real resistivity due to QPs becomes
\begin{equation}
    \rho = \frac{1}{\sigma} \approx \frac{\sigma_1}{\sigma_2^2}. 
\end{equation}
The resistance per unit length of a strip of superconductor with cross-section $A$ is
\begin{equation}
    R' = \frac{\rho}{A} = x_\mathrm{qp} \sqrt{\frac{2\Delta \omega}{\pi^2 \hbar}}L_\mathrm{k}',
    \label{eq:R'}
\end{equation}
where $L_\mathrm{k}' = (\omega\sigma_2 A)^{-1}$ is the kinetic inductance per unit length. 

\begin{figure}
    \centering
    \includegraphics[width=\linewidth]{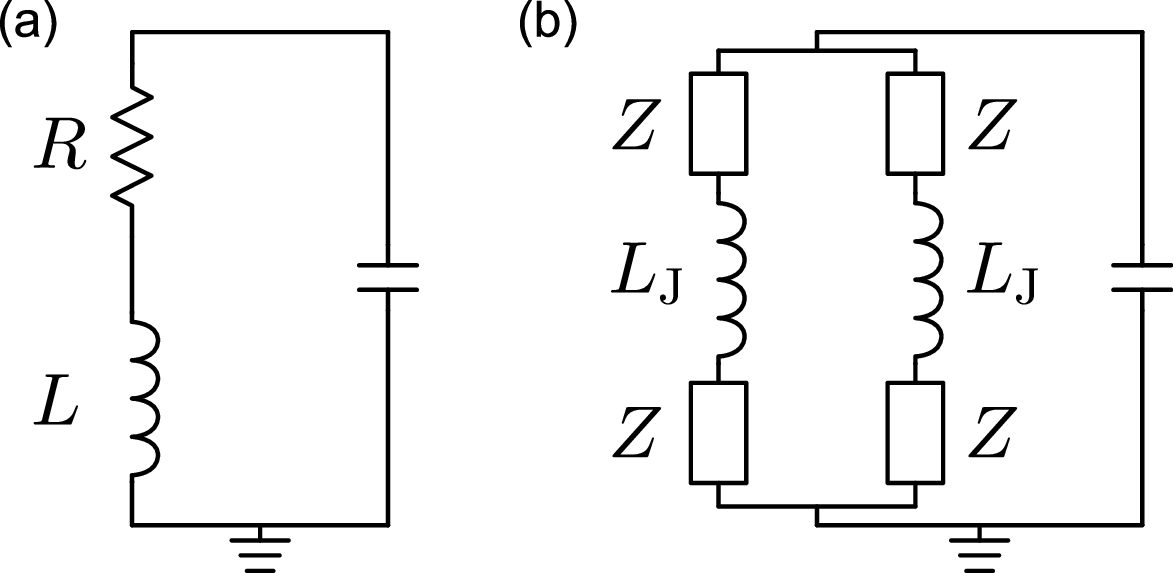}
    \caption{Electrical circuits used to model (a) the quarter-wave resonator and (b) the transmon. The impedances $Z$ represent the bandages connecting the JJ leads to the Nb ground plane and capacitor island of the transmon.}
    \label{fig:circuits}
\end{figure}

We model the quarter-wavelength resonator as a lumped RLC circuit with a resistor in series with an inductor and in parallel with a capacitor to ground, see \figpanel{fig:circuits}{a}. 
The series impedance $Z_\mathrm{s} = R + i\omega L$ can be converted to an equivalent parallel resistance
\begin{equation}
    R_\mathrm{p} = \left(\operatorname{Re}[1/Z_\mathrm{s}]\right)^{-1} = \frac{R^2 + (\omega L)^2}{R} \approx \frac{(\omega L)^2}{R},
\end{equation}
which we can insert into the standard expression $Q = R_\mathrm{p}/\omega L$ for the quality factor of a parallel RLC circuit to obtain the QP-limited quality factor
\begin{equation}
    Q_\mathrm{qp} = \frac{\omega L}{R} = \frac{\omega L'}{R'}, 
    \label{eq:Q_qp_def}
\end{equation}
where $L' = L_\mathrm{k}' + L_\mathrm{g}'$ and $L_\mathrm{g}'$ is the geometric inductance per unit length.
Inserting \eqref{eq:R'} into \eqref{eq:Q_qp_def} yields
\begin{equation}
    Q_\mathrm{qp} = \frac{L'}{L_\mathrm{k}'}\sqrt{\frac{\pi^2\hbar\omega}{2\Delta}} x_\mathrm{qp}^{-1}. 
\end{equation}

We can estimate $L_\mathrm{g}'$ from the measured resonance frequency $\omega$ and the length $l$ of the $Z=\SI{50}{\ohm}$ resonator if we assume $L_\mathrm{g}' \gg L_\mathrm{k}'$. Using $\omega/2\pi = \SI{6.7}{\giga\hertz}$ and $l=\SI{4.5}{\milli\meter}$, we get $L_\mathrm{g}' \approx \pi Z/(2\omega l) = \SI{400}{\nano\henry\per\meter}$. To estimate $L_\mathrm{k}'$ for the resonator, we can use \eqref{eq:sigma_2_simplified}:
\begin{equation}
    L_\mathrm{k}' = \frac{1}{\omega\sigma_2 A} \approx \frac{\hbar}{\pi\Delta\sigma_n A} = \frac{\hbar R_\mathrm{s}}{\pi\Delta w},
    \label{eq:L_k}
\end{equation}
where $R_\mathrm{s}$ is the normal-state sheet resistance of the Al film and $w = \SI{20}{\micro\meter}$ is the width of the center conductor. 
Note that \eqref{eq:L_k} only accounts for the current in the center conductor, not the current along the edges of the ground plane.
A better estimate can be obtained by substituting $1/w \rightarrow 1/w + 1/(w+2s)$ to account for additional currents in the ground plane, where $s = \SI{10}{\micro\meter}$ is the separation between the center conductor and the ground plane. 
We use resistivity measurements from Ref.~\cite{biznarovaMitigationInterfacialDielectric2024} of an Al film evaporated in the same tool under the same conditions and with the same thickness, which yields $L_\mathrm{k}' = \SI{10}{\nano\henry\per\meter}$ for our resonator. 
The calculated QP densities vs. optical power are presented in \tabref{tab:QP_density}.
\begin{table}
    \centering
    \setlength{\tabcolsep}{8pt}
    \begin{tabular}{lccc}
    \toprule
         \textbf{Optical power} & $\mathbf{Q}$ & $\mathbf{Q_\mathrm{qp}}$ & $\mathbf{x_\mathrm{qp}}$\\
         \SI{1}{\nano\watt} & $1.55 \times 10^6$ & $5.7 \times 10^6$ & $ 6\times 10^{-6}$ \\ 
         \SI{10}{\nano\watt} & $0.683 \times 10^6$ & $1.0 \times 10^6$ & $ 3\times 10^{-5}$ \\ 
         \SI{100}{\nano\watt} & $0.177 \times 10^6$ & $0.19 \times 10^6$ & $ 2\times 10^{-4}$ \\ 
    \end{tabular}
    \caption{Estimated QP density in the Al resonator as a function of applied optical power and corresponding measured internal quality factor $Q$. 
    For zero optical power we have $Q = 2.1\times10^6$, and we assume QPs to be the only source of excess energy loss relative to this value for $P_\mathrm{o} \geq \SI{1}{\nano\watt}$.}
    \label{tab:QP_density}
\end{table}

\section{Quasiparticle density in the Nb transmon's bandages} 
\label{sec:QP_density_q}
The Nb transmon is designed with low-gap Al bandages that will trap QPs. 
We want to find the impact of a finite Ohmic resistance in the bandages on the transmon's quality factor. 
Similarly to the resonator discussed above, we can then derive an estimate for the QP density in the bandages required to reproduce the measured quality factor of the transmon. 
We assume that the QP density in the Nb and the junction leads is negligible compared to the bandages. 
We also assume the gap engineering of the JJs prevents all QP tunneling across the junctions in the SQUID. 
We discuss the possibility of hot QPs tunneling through the JJ in Appendix~\ref{sec:hot_QPs}.

Our model of the transmon is shown in \figpanel{fig:circuits}{b}. 
Each JJ with inductance $L_\mathrm{J}$ in the SQUID of the flux-tunable transmon is connected to the Nb base layer with Al bandages.
Each bandage has impedance $Z = R + i\omega L$ with Ohmic resistance $R$ caused by the QPs. 
The total impedance of the SQUID consisting of four bandages and two JJs is equivalent to a single resistance $R$ in series with an inductance $L + L_\mathrm{J}/2$. 
This is now an RLC circuit identical to the one shown in \figpanel{fig:circuits}{a}. 
Following the derivation of the quality factor for this circuit above, we get
\begin{equation}
    Q_\mathrm{qp}^\mathrm{transmon} \approx \frac{L_\mathrm{J}}{2L_\mathrm{k}}\sqrt{\frac{\pi^2\hbar\omega}{2\Delta}} x_\mathrm{qp}^{-1},
\end{equation}
where we used $L_\mathrm{J} \sim \SI{10}{\nano\henry} \gg L$. 
We estimate the kinetic inductance of the bandages to be $L_k \sim \SI{0.5}{\pico\henry}$ using the same material properties as the Al film forming the resonator. 
This yields
\begin{equation}
    \frac{Q_\mathrm{qp}^\mathrm{transmon}}{Q_\mathrm{qp}^\mathrm{resonator}} \sim 500.
\end{equation}
We therefore conclude that, according to this crude model, the QP density in the bandages must be 500 times greater than in the bulk Al film to cause the transmon and the resonator to have similar quality factors. 

\section{Resistance simulations}
\label{sec:simulations}
\begin{figure}
    \centering
    \includegraphics[width=\linewidth]{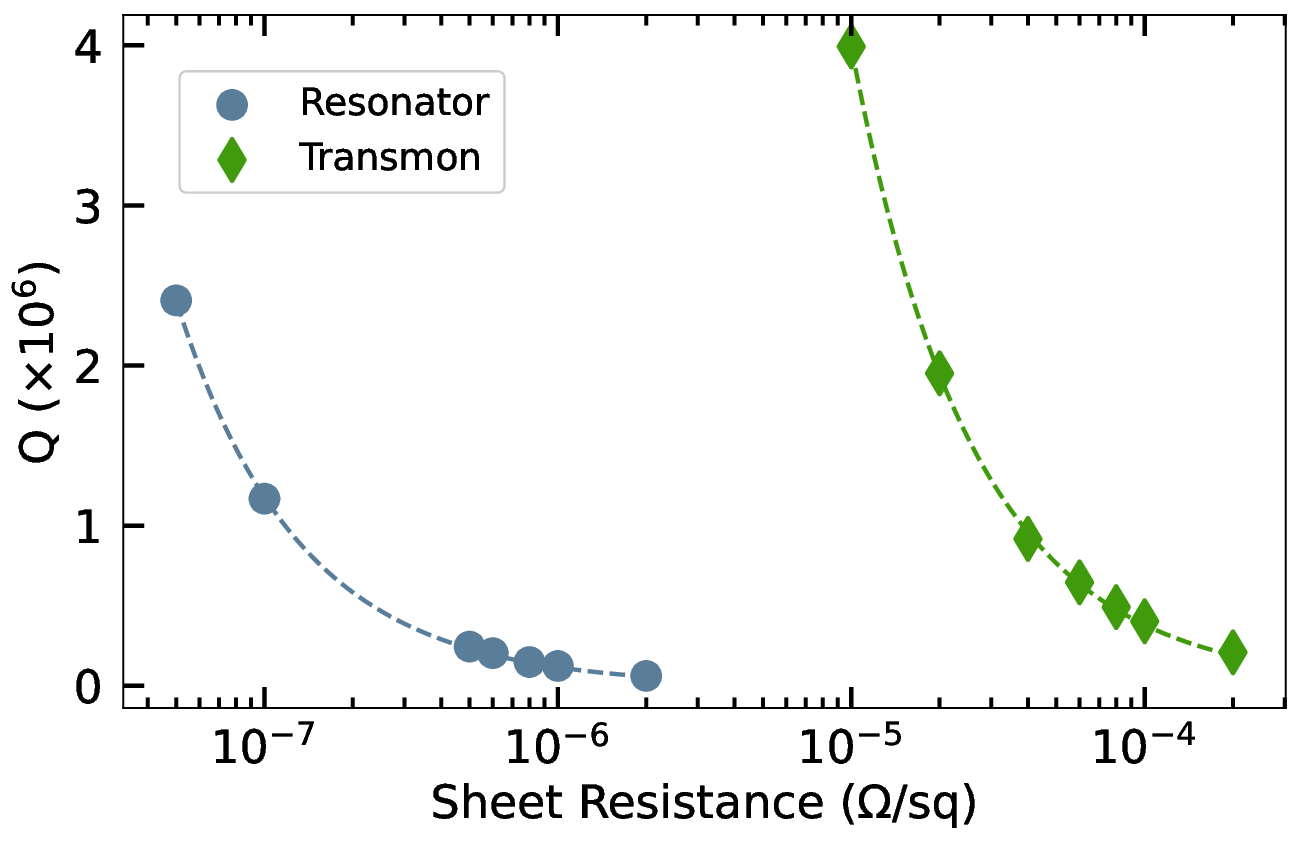}
    \caption{Simulated quality factor vs. sheet resistance. Dashed lines are fits to $Q = (\alpha R_\mathrm{s} + \beta)^{-1}$, where $\alpha, \beta$ are fitting parameters.}
    \label{fig:resistance_sim}
\end{figure}

We perform eigenmode simulations of the standalone resonator and the transmon in ANSYS HFSS to further investigate the effects of QP poisoning of the bandages in the transmon. 
We model the superconductor as a sheet with finite resistance due to excess QPs that dissipate microwave power. 
Because the superconducting gap is uniform across the resonator and the surrounding ground plane, we assume a uniform distribution of quasiparticles and hence a uniform sheet resistance. 
We then vary the sheet resistance and track the calculated internal quality factor of the resonator. 
We are only interested in the quality factor of the resonator and neglect the change in inductance caused by the QPs, which induces a small shift ($\sim \SI{30}{\kilo\hertz})$ in the resonator frequency when the quality factor of the resonator has dropped by an order of magnitude relative to the $P_\mathrm{o}=0$ baseline. 
As expected, we observe an inverse relationship between the internal quality factor and the sheet resistance, as shown in \figref{fig:resistance_sim}. 
The required sheet resistance $R_\mathrm{s}$ to reproduce the measured QP-limited internal quality factor at $P_\mathrm{o} = \SI{100}{\nano\watt}$ is $R_\mathrm{s}\simeq \SI{0.7}{\micro\ohm\per\sq}$. 

Next, we want to estimate how high the resistance in the bandages of the Nb transmon must be to limit its quality factor. 
We build a model of the transmon where the Nb base layer is represented by a perfect conductor. 
The leads of the JJs are given a resistance of $R_\mathrm{s} =\SI{1}{\nano\ohm\per\sq}$, which is low enough to not limit the quality factor as we sweep the resistance of the bandages. 
The resulting quality factor for the transmon is shown in \figref{fig:resistance_sim}.
The combination of relatively weak electric field and small size of the bandages allows the transmon's quality factor to remain high despite an elevated sheet resistance. 
The simulation results show that the transmon's quality factor is not limited by bandage resistance below $R_\mathrm{s} \sim \SI{10}{\micro\ohm\per\sq}$.
We find that the required resistance in the bandages to reproduce the measured quality factor in the transmon at $P_\mathrm{o} = \SI{100}{\nano\watt}$ is $\sim400$ times higher than what was observed in the Al resonator at the same optical power, which is in reasonable agreement with the lumped circuit model. 

\begin{figure*}
    \centering
    \includegraphics[width=\linewidth]{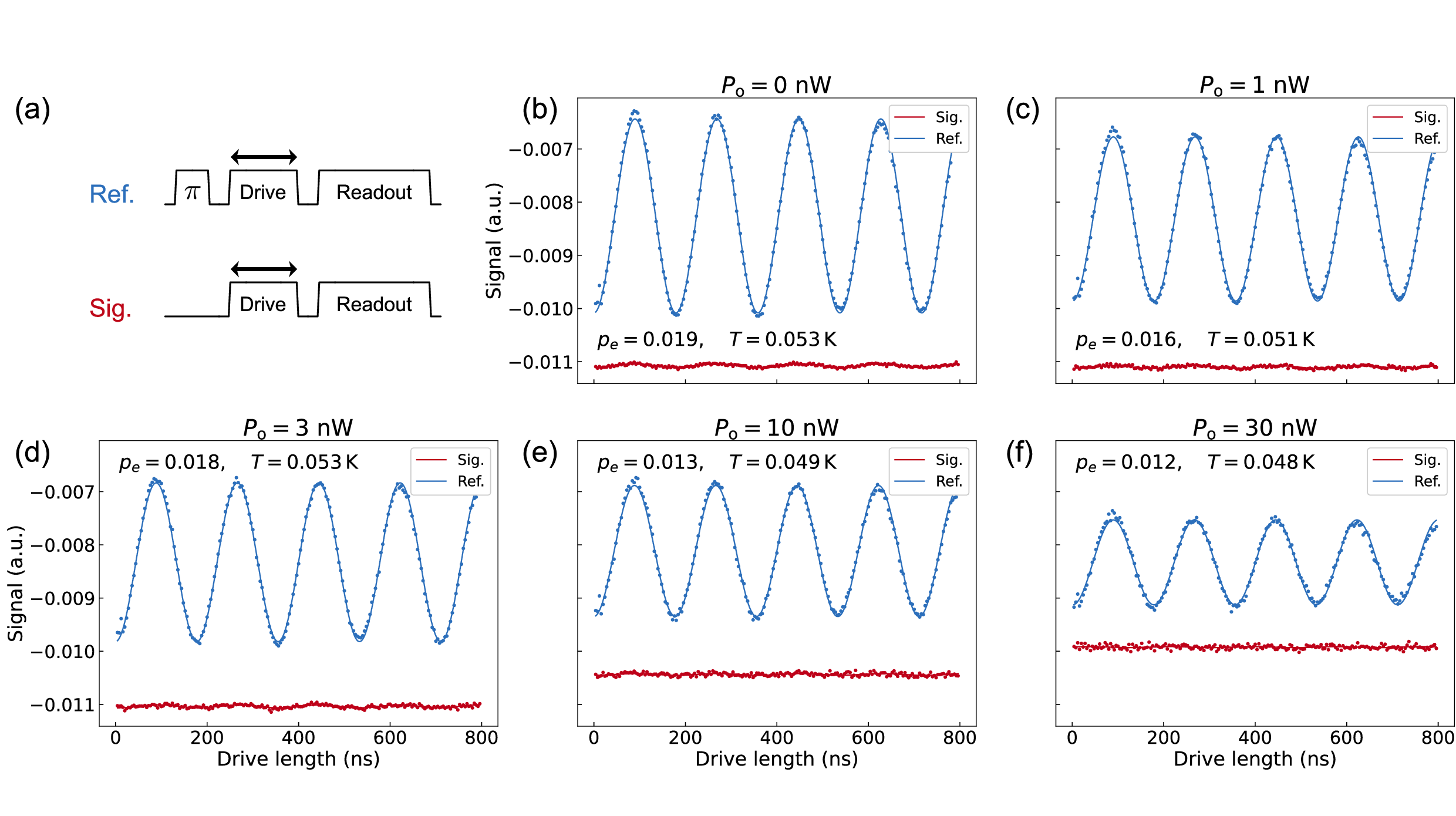}
    \caption{Transmon thermometry under optical illumination. (a) The pulses used to extract the $\vert e\rangle$ population. (b)-(f) The resulting Rabi oscillations from which we estimate the population and the corresponding effective temperature of the transmon.}
    \label{fig:thermometry}
\end{figure*}

\section{Hot quasiparticles}
\label{sec:hot_QPs}
Previous studies of QPs have found that, although the QP density is not in thermal equilibrium with the environment, the energy distribution can be in thermal equilibrium with the phonon bath of the chip~\cite{serniakHotNonequilibriumQuasiparticles2018, connollyCoexistenceNonequilibriumDensity2024, nhoRecoveryDynamicsGapEngineered2026}.
Under optical illumination, the Nb film covering the chip absorbs photons that break Cooper pairs and create QPs. 
When the QPs recombine in the Nb film, they emit phonons at frequencies $\geq 2\Delta_\mathrm{Nb}/h$. 
The phonons can either break more Cooper pairs or escape to the Si substrate.
Under strong optical illumination, we therefore expect that the Si substrate hosts a phonon bath that is not in equilibrium with the mixing chamber of the dilution refrigerator. 
The QPs in the leads of the JJs can then couple to this phonon bath.
This coupling may produce a QP energy distribution characterized by an effective temperature higher than that of the mixing chamber. 
The additional thermal energy could allow QPs to absorb energy from the transmon and tunnel across the JJ despite the gap difference. 

If QPs are in thermal equilibrium with a hot phonon bath kept at an elevated temperature $T$ by constant emission of high-energy phonons from the Nb film, we expect the tunneling rate of QPs through the junction to follow an Arrhenius-type activation~\cite{nhoRecoveryDynamicsGapEngineered2026}, 
\begin{equation}
    \Gamma_\mathrm{tunnel} \propto \exp\left(-\frac{\delta\Delta - \hbar\omega_\mathrm{q}}{k_\mathrm{B} T}\right) x_\mathrm{qp},
\end{equation}
where $\delta\Delta$ is the gap difference across the junction and $\omega_\mathrm{q}$ is the qubit frequency. 
Given that $\delta\Delta > \omega_\mathrm{q} \gg k_\mathrm{B}T/\hbar$, the induced qubit relaxation rate from QP tunneling is likely to be almost identical to the tunneling rate. 
The QP density $x_\mathrm{qp}$ and the temperature $T$ both depend on the applied optical power. 
However, the absence of a strong frequency shift in the transmon qubit suggests that $x_\mathrm{qp}$ depends only weakly on optical power below $P_\mathrm{o} = \SI{10}{\nano\watt}$. 

Constructing a model of the hot phonon bath in the Si chip and its coupling to the cold fridge bath and the QPs in the Nb film to find $T (P_\mathrm{o})$ is beyond the scope of this work. 
As we show in Appendix~\ref{sec:thermometry}, if such a hot phonon bath is present, it does not couple directly to the transmon.

\section{Transmon thermometry}
\label{sec:thermometry}
We measure the population in $\vert e\rangle$ after a passive reset using the protocol from Ref.~\cite{geerlingsDemonstratingDrivenReset2013}.
The pulse sequence used in the protocol is illustrated in \figpanel{fig:thermometry}{a}.
For the reference, we first invert the $\vert g\rangle$ and $\vert e\rangle$ populations with a $\pi$-pulse. 
The transmon is then driven by a pulse centered on the $\vert e\rangle \rightarrow \vert f\rangle$ transition for a variable amount of time before readout. 
The drive pulse produces Rabi oscillations with amplitude $A$ proportional to the population in the state $\vert e\rangle$. 
By comparing the amplitudes $A_\mathrm{sig}, A_\mathrm{ref}$ of the Rabi oscillations depending on whether a population inversion preceded the drive, we can estimate the thermal population as $p_e \approx A_\mathrm{sig}/(A_\mathrm{sig} + A_\mathrm{ref})$. 
The resulting Rabi oscillations measured under increasing levels of optical power at the chip are shown in \figpanels{fig:thermometry}{b}{f}.
We observe no increase in the transmon's excited-state population as we increase the optical power. 
The amplitude of the Rabi oscillations decreases with increasing optical power because our ability to discriminate the states $\vert e\rangle$ and $\vert f\rangle$ deteriorates, as we show in the main text. 
The amplitude of the reference signal does not decay appreciably for the longest pulse lengths of $\SI{800}{\nano\second}$ even at $P_\mathrm{o} = \SI{30}{\nano\watt}$, which highlights the resilience of the transmon to QP-induced decay. 

\section{Pulsed optics}
Previous studies~\cite{bankerPhotonicPhononicBand2022,benevidesQuasiparticleDynamicsSuperconducting2024} of transmon qubits subjected to optical laser pulses have found that the relaxation rate of the transmon increases after a pulse.
The relaxation rate remains high for more than $\SI{10}{\micro\second}$ after the pulse due to slow QP relaxation and trapping in the JJs. 
Analogously, superconducting qubits have been found to recover on a similar timescale after a high-energy impact event and corresponding error burst. 

\begin{figure}
    \centering
    \includegraphics[width=\linewidth]{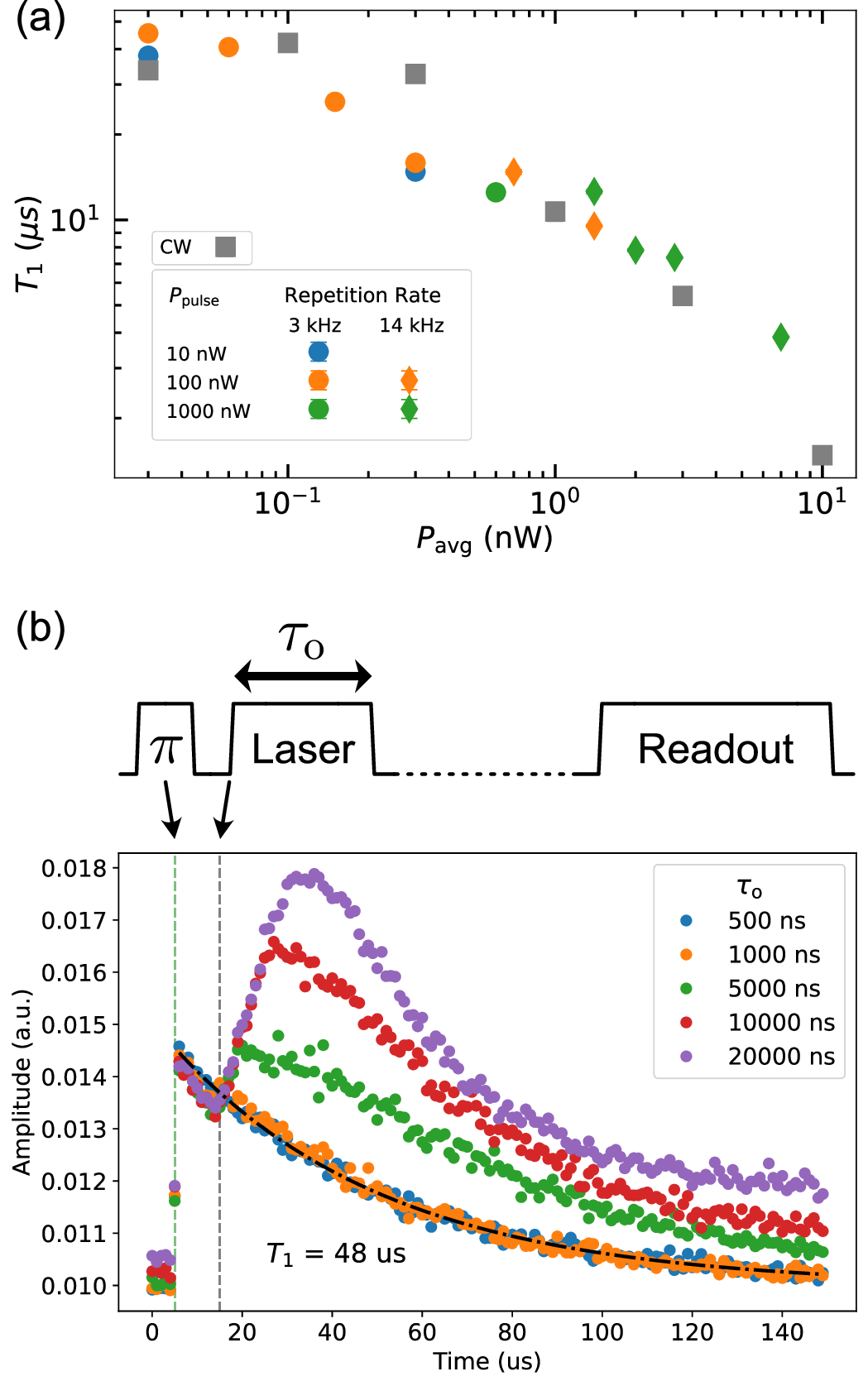}
    \caption{Transmon response to pulsed optical laser light. (a) $T_1$ of the Al transmon vs average optical power $P_\mathrm{avg}$. (b) Pulse sequence (top) and readout trace (bottom) of the Nb transmon for a laser pulse with $P_\mathrm{pulse} = \SI{1}{\micro\watt}$ and $R = \SI{1}{\kilo\hertz}$.}
    \label{fig:pulsed_optics}
\end{figure}

We first measure the relaxation rate of the Al transmon when we apply laser pulses to the chip. 
As expected, the measured $T_1$ after a laser pulse decreases as we increase the pulse power $P_\mathrm{pulse}$.
We observe the same effect when we increase the pulse length $\tau_\mathrm{o}$ and the repetition rate $R$. 
As can be seen in \figpanel{fig:pulsed_optics}{a}, the measured $T_1$ depends on the average optical power, 
\begin{equation}
    P_\mathrm{avg} = P_\mathrm{pulse}\times \tau_\mathrm{o} \times R. 
\end{equation}
There is no clear sign that the transmon is more sensitive to $P_\mathrm{pulse}$, $\tau_\mathrm{o}$, or $R$, consistent with the findings of Ref.~\cite{benevidesQuasiparticleDynamicsSuperconducting2024}, which studied how a transmon responds as $P_\mathrm{pulse}$ and $\tau_\mathrm{o}$ are varied. 
The data point to a slow recovery of the Al transmon qubit following a laser pulse.  

Next, we study the relaxation of the Nb transmon under pulsed optical illumination. 
The pulse sequence and example of a resulting readout signal are shown in \figpanel{fig:pulsed_optics}{b}.
We are unable to find any combination of $P_\mathrm{pulse}$, $\tau_\mathrm{o}$, and $R$ that reveals an increased relaxation rate in the readout trace.
At values of $P_\mathrm{avg}$ for which we begin to observe a degradation of the Al transmon, the Nb transmon remains insensitive to the optical pulse. 
However, as we increase the optical power further, the readout signal suggests that the transmon is leaving the qubit manifold during readout. 
The qubit becomes increasingly ionized as we increase the duration of the optical pulse. 
Once the optical light is switched off, the transmon initially recovers on a timescale of $\sim\SI{100}{\micro\second}$.
For the longest optical pulses in our experiment, we also observe a residual ionization with a slower recovery rate. 
At $P_\mathrm{avg} \sim \SI{10}{\nano\watt}$, the transmon can no longer fully recover in $\SI{1}{\milli\second}$. 

\section{Maximum optical power and compatibility with optically driven devices}
While the optical powers in our study are low compared to what is desired in typical experiments involving optically driven devices such as microwave-to-optical transducers, our results nonetheless show promise for gap-engineered qubits to be directly connected to such devices. 
Our packaging is essentially an optical cavity designed to confine the light and cause uniform QP poisoning across the microwave chip.
This is a worst-case scenario for a transducer coupled to a transmon qubit.
The optical-power thresholds reported here are specific to the deliberately reflective package, which is intended to facilitate a direct comparison between the Al and Nb/Au devices without precise positioning of the optical fiber. 

In a separate experiment, we coated the inside of the copper package with Acktar FractalBlack, a material that absorbs light at $\SI{1550}{\nano\meter}$.
We observed that the optical power required to raise the relaxation rate of the transmon qubit was about 100 times higher than for the reflective copper package used in this study.  
Unfortunately, the readout efficiency of the transmon on the chip mounted in the package with the absorbing material was too low to efficiently characterize the qubit response beyond a proof-of-concept measurement. 
The cause of the poor readout fidelity remains unclear, but was likely due to losses on the output line, not the light-absorbing material. 
We leave the investigation of the maximum optical power that a gap-engineered transmon qubit can withstand in an optimized package to future work.

\bibliography{Qubit_paper}
\end{document}